\documentclass{elsart}

 \usepackage{graphicx}
\usepackage{amssymb}

\begin{document}

\begin{frontmatter}

% Title, authors and addresses

% use the thanksref command within \title, \author or \address for footnotes;
% use the corauthref command within \author for corresponding author footnotes;
% use the ead command for the email address,
% and the form \ead[url] for the home page:
% \title{Title\thanksref{label1}}
% \thanks[label1]{}
% \author{Name\corauthref{cor1}\thanksref{label2}}
% \ead{email address}
% \ead[url]{home page}
% \thanks[label2]{}
% \corauth[cor1]{}
% \address{Address\thanksref{label3}}
% \thanks[label3]{}

\title{New Results for Double--Beta Decay of $^{100}$Mo to Excited Final States of $^{100}$Ru Using the TUNL-ITEP Apparatus}

% use optional labels to link authors explicitly to addresses:
% \author[label1,label2]{}
% \address[label1]{}
% \address[label2]{}

\author[Duke]{M.F. Kidd}
\author[Duke]{J.H. Esterline}
\author[Duke]{W. Tornow}
\ead{tornow@tunl.duke.edu}
\author[ITEP]{A.S. Barabash}
\author[ITEP]{V.I. Umatov}

\address[Duke]{Department of Physics, Duke University and Triangle Universities Nuclear Laboratory, Durham, North Carolina 27708--0308, USA}
\address[ITEP]{Institute for Theoretical and Experimental Physics, 117259 Moscow, Russia}

\begin{abstract}
% Text of abstract
The coincidence detection efficiency of the TUNL--ITEP apparatus designed for measuring half-life times of two-neutrino double-beta (2$\nu\beta\beta$) decay transitions to excited final states in daughter nuclei has been measured with a factor of 2.4 improved accuracy. In addition, the previous measuring time of 455 days for the study of the $^{100}$Mo 2$\nu\beta\beta$ decay to the first excited 0$_1^+$ state in $^{100}$Ru has been increased by 450 days, and a new result (combined with the previous measurement obtained with the same apparatus) for this transition is presented: T$_{1/2}$= [5.5$^{+1.2}_{-0.8}$ (stat) $\pm$ 0.3 (syst)]$\times$10$^{20}$ y.  Measured 2$\nu\beta\beta$ decay half-life times to excited states can be used to test the reliability of nuclear matrix element calculations needed for determining the effective neutrino mass from zero-neutrino double-beta decay data. We also present new limits for transitions to higher excited states in $^{100}$Ru which, if improved, may be of interest for more exotic conjectures, like a bosonic component to neutrino statistics.
\end{abstract}

\end{frontmatter}

% main text
%\section{}
%\label{}
\section{INTRODUCTION}

The half-life time of the zero-neutrino double-beta (0$\nu\beta\beta$) decay is related to the neutrino mass by the equation
 \begin{equation}
 [T_{1/2}^{0\nu}]^{-1}=G_{0\nu}(M_F^{0\nu}-(g_A/g_V)^2M_{GT}^{0\nu})^2<\!m_\nu\!>^2
 \label{eqno1}.
 \end{equation}

Here, G$_{0\nu}$ is the two-body phase-space factor which includes the coupling constant, $<$m$_\nu\!>$ is the effective Majorana electron neutrino mass, and M$_F^{0\nu}$ and M$_{GT}^{0\nu}$ are the Fermi and Gamow-Teller nuclear matrix elements, respectively. Therefore, extracting the neutrino mass from the measured half-life time requires the knowledge of the 0$\nu\beta\beta$ nuclear matrix elements. Information on the nuclear matrix elements is currently obtained from quasi-particle random phase approximation (QRPA) and nuclear shell-model calculations \cite{Fae}. The parameters in these calculations, in particular the strength of the particle-particle interaction g$_{pp}$, in turn are adjusted using measured half-life times of single-beta, and more recently, two-neutrino double-beta (2$\nu\beta\beta$) decay data. Because the behavior of g$_{pp}$ is different for the transition to the ground state compared to transitions to excited states in the daughter nucleus, measuring 2$\nu\beta\beta$ decay to excited final states provides an excellent test of the theoretical calculations \cite{Aun}.  Furthermore, a very interesting possibility is the search for the 0$\nu\beta\beta$ transition to the 0$^+_1$ excited state of the daughter nucleus.  This transition provides a clear-cut signature:  in addition to the two electrons with a fixed total energy, there are two gamma rays whose energies are strictly fixed as well.  In a hypothetical experiment detecting all decay products with high efficiency and sufficient energy and spatial resolution, the background can be reduced to nearly zero.  It is quite possible this idea will be used in future large scale 0$\nu\beta\beta$ searches such as MAJORANA \cite{Maj}, \cite{Avi08}, GERDA \cite{Abt}, \cite{Sim08}, CUORE \cite{Arn04}, \cite{Ban08}, or SuperNEMO \cite{Bar02},\cite{Ohs08}.  In addition, in Ref. \cite{Sim02} it was stated that the detection of the excited state transition will provide the opportunity of distinguishing between different 0$\nu\beta\beta$ mechanisms, such as light and heavy Majorana neutrino exchange or trilinear R-parity breaking mechanisms.

Recently it was speculated \cite{Dol05} that neutrinos may violate the Pauli exclusion principle (PEP) and therefore, at least partly obey Bose-Einstein statistics (see also \cite{Cuc96}). As a consequence, neutrinos could form a Bose condensate which may account for parts or even all of the dark matter in the universe. As discussed in \cite{Bar07} the possible violation of the PEP has interesting  consequences for 2$\nu\beta\beta$ decay.  It not only modifies the energy and angular distributions of the emitted electrons, but it also strongly affects the 2$\nu\beta\beta$ decay rates to ground and excited states in daughter nuclei. Following \cite{Bar07}, the half-life time ratios for transitions to excited 2$^+$ states and the 0$^+$ ground state are by far the most sensitive way of obtaining bounds on the conjecture of a substantial bosonic component to neutrino statistics \cite{Dol05}.  As a result, information on the decay rates to excited states is needed to test this new and potentially far-reaching hypothesis.

Presently, 2$\nu\beta\beta$ decay has been observed for ten nuclei \cite{Barrev}, but data for the 2$\nu\beta\beta$ decay to excited final states exist only for $^{100}$Mo \cite{Bar95}--\cite{Arn} and $^{150}$Nd \cite{Bar04}. 

Recently, a new determination of the half-life time of $^{100}$Mo for the transition to the first excited 0$_1^+$ state in $^{100}$Ru was reported by Hornish {\it et al.} \cite{Hor}. In addition, limits to higher excited states were also given in \cite{Hor}. These measurements were performed on ground level at TUNL using the coincidence technique. In the case of $^{100}$Mo, the excited 0$_1^+$ state of the daughter nucleus $^{100}$Ru de-excites via the decay sequence 0$_1^+\rightarrow 2_1^+\rightarrow 0_{gs}^+$, emitting two $\gamma$ rays in coincidence with energies of E$_{\gamma1}$=590.8 keV and E$_{\gamma2}$=539.5 keV.

The TUNL-ITEP $\beta\beta$ decay setup consists of two cylindrical high-purity germanium (HPGe) detectors with the $\beta\beta$ decay source sandwiched between them. The HPGe detectors are surrounded by a NaI(Tl) annulus which in turn is placed inside of a lead-brick enclosure. The combination of active and passive shielding suppresses the coincidence background to such an extent that successful measurements on ground level are feasible, provided  $\beta\beta$ decay source masses of about 1 kg are available and half-life times do not exceed 10$^{22}$ years. The $\beta\beta$ decay source used at TUNL was a metallic disk with mass of 1.05 kg, enriched to 98.4\% in $^{100}$Mo. After counting for 455 days, 22 coincidence events were detected with a continuous background estimated to be 2.5 events. Using the measured efficiency of the detector system, a half-life time of T$_{1/2}$=[6.0$^{+1.9}_{-1.1}$(stat)$\pm$0.6(syst)]$\times$10$^{20}$ y was reported in \cite{Hor}.      

Here we report an improved measurement of the coincidence efficiency of the TUNL-ITEP $\beta\beta$ decay apparatus, and new data taken for $^{100}$Mo during 450 days of counting.

\section{COINCIDENCE-EFFICIENCY MEASUREMENT}

We closely followed the procedure described in detail in \cite{Hor}. The radioactive source used to measure the efficiency was $^{102}$Rh, which was produced at TUNL via the reaction $^{102}$Ru(p,n)$^{102}$Rh initiated with 5 MeV incident protons on a natural ruthenium target (31.6\% $^{102}$Ru). This particular source was chosen for several reasons. First, the  0$_1^+\rightarrow 2_1^+\rightarrow 0_{gs}^+$ decay sequence not only results in $\gamma$ rays in coincidence, but also mimics the decay scheme of the excited final 0$_1^+$ state of $^{100}$Mo. Therefore, corrections for the angular distribution are unnecessary. The associated $\gamma$ ray--energies, 468.6 keV and 475.1 keV, are also close to those of $^{100}$Ru (539.5 keV and 590.8 keV), the daughter nucleus of $^{100}$Mo. Finally, the 205.1 $\pm$ 1.9 day half-life of $^{102}$Rh is long enough to allow shorter lived contaminants to decay away prior to measurement. The activity of the source was determined by comparing the intensity of the 475.1 keV $\gamma$--ray peak to that of the 661.6 keV $\gamma$--ray peak from a calibrated $^{137}$Cs source.

The HPGe detectors of \cite{Hor} were 8.8 cm in diameter and 5.0 cm in thickness. Because the $^{100}$Mo source used for the measurement of the half-life time in \cite{Hor} and in our work described in Sec. 3 was in form of a 10.6 cm diameter and 1.1 cm thick disk, the radial dependence of the 468.6 keV -– 475.1 keV coincidence efficiency across the detector faces was crucial. We sandwiched our 1 $\times$ 1 $\times$ 0.5 mm$^3$ ruthenium/rhodium source (in the following referred to as $^{102}$Rh source) between ten disks (five on each side) of 10 cm diameter and 0.1 cm thickness of natural molybdenum to ensure that the overall thickness was the same as in the original 2$\nu\beta\beta$ decay experiment on $^{100}$Mo. This procedure yields the same average $\gamma$--ray attenuation provided the energy difference between the $\gamma$--ray pairs (468.6 keV -– 475.1 keV versus 539.5 keV -– 590.8 keV) is properly taken into account. Furthermore, it enables the $^{102}$Rh source to be positioned not only at different radii, but also at different locations along the {\it z} axis, {\it i.e.}, along the common axis of the two HPGe detectors.

\begin{figure}
\begin{center}
 \includegraphics[width = 5.25in]{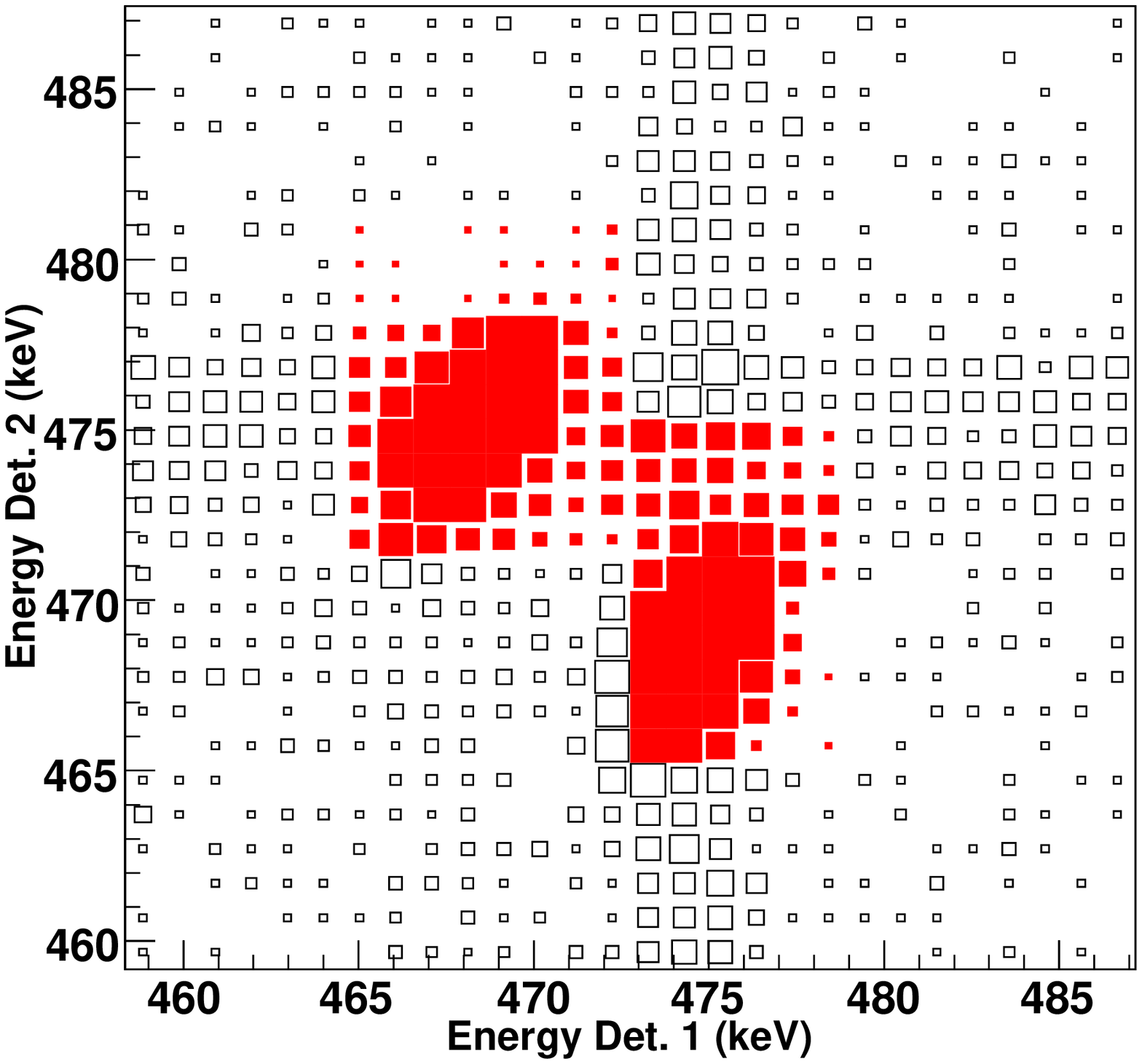}%
 \caption{Pulse height of HPGe detector 2 versus pulse height of HPGe detector 1 with the events associated with the 468.6 keV -– 475.1 keV decay sequence of $^{102}$Ru shaded in.\label{fig:2Denergy}}
\end{center}
 \end{figure}

The previous coincidence efficiency measurement \cite{Hor} was made along the radius $r$ on the front faces of the HPGe detectors at the four locations $r$=0, 2, 4 and 5 cm, {\it i.e.}, in only one direction, assuming cylindrical symmetry of the detectors. The current measurements were performed across the entire diameter and in two perpendicular directions, {\it i.e.}, horizontally and vertically. Near the center we used 0.5 cm increments to better determine the efficiency in that region. A total of 25 data points were obtained in this manner between -4.5 cm $\leq$ {\it r} $\leq$ 4.5 cm with the $^{102}$Rh source sandwiched between two 0.5 cm--thick molybdenum disks.  The measuring time for each data point varied between 12 and 48 hours, always obtaining at least 8000 coincidence events. At each location of the $^{102}$Rh source the number of 468.6 keV -- 475.1 keV coincidences was determined from the two--dimensional data area of pulse height in HPGe detector 1 versus pulse height in HPGe detector 2 (see Fig. \ref{fig:2Denergy}).

\begin{figure}
\begin{center}
 \includegraphics[width = 5.25in]{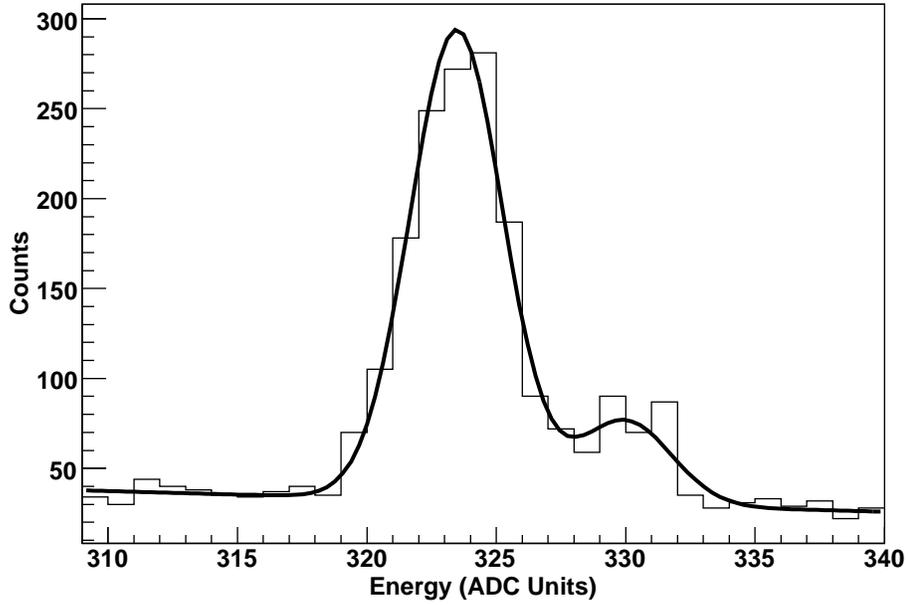}%
 \caption{An example of a projection of the pulse height distribution from Fig. \ref{fig:2Denergy} onto the HPGe detector 2 axis and fit to data.\label{fig:proj}}
\end{center}
 \end{figure}

The data were analyzed using the ROOT C/C++ Interpreter. The program used gates and projections (see Fig. \ref{fig:proj}) along the axes to set close gates around the coincidence peaks and, after background subtraction, obtained the yields in each peak. The time normalized yields were then corrected for dead time ($<$3\%) and normalized to the activity of the $^{102}$Rh source, taking into account its half-life. We measured this half-life independently and determined it to be 203.3 $\pm$ 2.8 days.  Combined with an existing world average of 206.7 $\pm$ 2.6 days from \cite{41Mi02}--\cite{61Mc10}, we obtained a half life of 205.1 $\pm$ 1.9 days.  In order to
obtain the efficiency for the $^{100}$Mo 2$\nu\beta\beta$ decay to the 0$_1^+$ state of $^{100}$Ru the yields had to be corrected not only for the difference in attenuation between the $^{102}$Rh and $^{100}$Ru $\gamma$--ray lines of interest, but also for the energy dependent detection efficiency of our two identical HPGe detectors. For that purpose, the relative detection efficiency was measured with a $^{152}$Eu source.

\begin{figure}
\begin{center}
 \includegraphics[width = 5.25in]{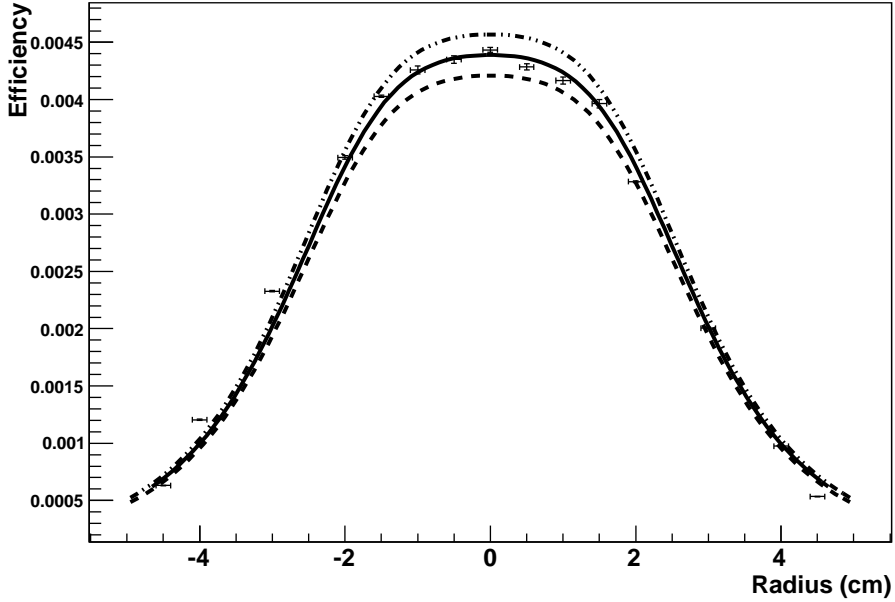}%
 \caption{Average of coincidence detection efficiency data obtained for the horizontal and vertical scans as a function of $r$ for the E$_{\gamma1}$=590.8 keV and E$_{\gamma2}$=539.5 keV coincidence. The data were taken with the$^{102}$Rh source (E$_{\gamma1}$=468.6 keV and E$_{\gamma2}$=475.1 keV) and then corrected for detection efficiency and attenuation differences between the two $\gamma$--ray pairs involved.  The curve through the data presents a least--squares fit. The upper and lower curves indicate the $\pm$5.1\% scale uncertainty associated with our data.\label{fig:cavg}}
\end{center}
 \end{figure}

As has been expected, the radial dependence of the coincidence efficiency was almost identical for the horizontal and vertical scans, within error associated with the exact location of the $^{102}$Rh source. Figure \ref{fig:cavg} shows the average of the horizontal and vertical efficiencies as a function of {\it r}. The error bars represent the statistical uncertainty of the data. Here, the effect of the position uncertainty of the $^{102}$Rh source is added in quadrature to the statistical uncertainty. The coincidence efficiency is rather small, close to 0.4\% in the -1 cm $<$ {\it r} $<$ 1 cm range and then dropping smoothly to about 0.1\% at {\it r}=4 cm. As a result, only about 3\% of our $^{100}$Mo disk provides 2$\nu\beta\beta$ decay related $\gamma$ rays with the maximum detection efficiency. However, about 30\% of the disk provides $\gamma$ rays with an efficiency of 50\% of the maximum efficiency obtained at $r$=0 cm. 

The curve through the data points in Fig. \ref{fig:cavg} is a least-square fit using the functional form
 \begin{equation}
 \epsilon_{\gamma\gamma}(r)=\frac{a}{1+br^2+cr^4}  
 \label{eqno2},
 \end{equation}

where $a$, $b$, and $c$ are free parameters and {\it r}=0 refers to the center on the front face of the HPGe detectors. A careful inspection reveals a slight asymmetry in the coincidence efficiency, providing slightly larger values for {\it r}$<$ -2 cm. The lower (dashed) and upper (dashed-dotted) curves shown in Fig. \ref{fig:cavg} represent our systematic uncertainty of 5.1\%, including the $\pm$3\% scale uncertainty associated with the $^{137}$Cs $\gamma$-ray source.

\begin{figure}
\begin{center}
 \includegraphics[width = 5.25in]{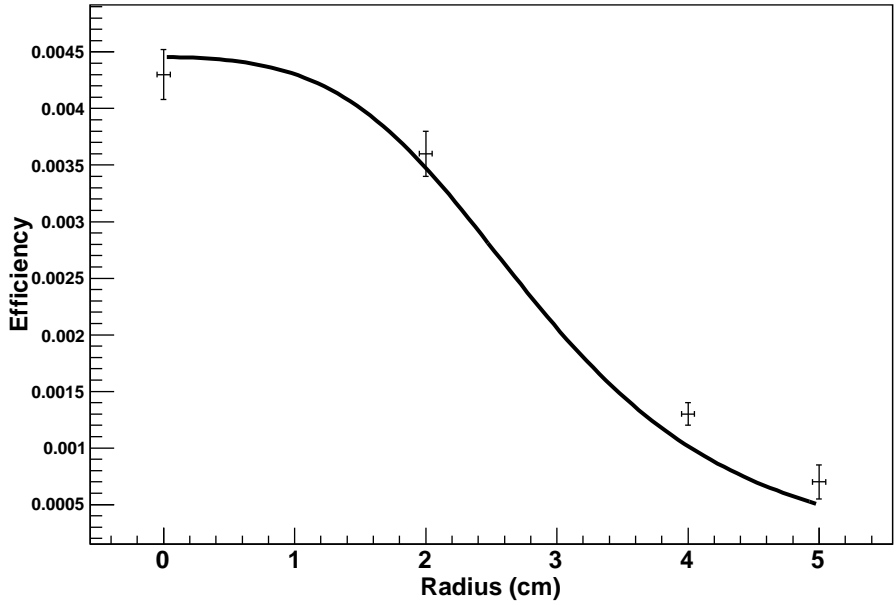}%
 \caption{Comparison of previous coincidence efficiency data \cite{Hor} and fit to present data (solid curve).\label{fig:effcomp}}
\end{center}
 \end{figure}

The radial contribution to the coincidence efficiency was calculated using

 \begin{equation}
 \epsilon_{r}=\frac{2\pi\int\epsilon_{\gamma\gamma}(r)rdr}{2\pi\int rdr},
 \label{eqno3}
 \end{equation}

where $\epsilon_{\gamma\gamma}(r)$ is the best fit obtained from an asymmetric fit to the data. This value is then corrected by the {\it z}-dependence of the coincidence efficiency measured in the present work and confirmed by Monte-Carlo simulation, where {\it z} is the distance from the face of the detector.  There is a 10\% decrease in efficiency at the center of the 1 cm thick molybdenum disk compared to the front and back faces.  The resulting efficiency value is $\epsilon_{tot}$=(0.182$\pm$0.009)\%, where the total uncertainty of 5.1\% is due to the contributions listed in Table \ref{tab:error}.  They include an estimated 3\% uncertainty due to the slightly irregular shape of our $^{100}$Mo disk.  The previous measurements described in \cite{Hor} provided $\epsilon_{tot}$=(0.219$\pm$0.022)\%. The associated data are shown in Fig. \ref{fig:effcomp} in comparison with a fit to the present data (solid curve). The main reason for the difference between the previous and the present value is due to an improved determination of the {\it z}-dependence of the efficiency.

%In addition, the {\it z} dependence of the coincidence efficiency was measured at {\it r}=0 cm for -0.5 cm$\leq${\it z}$\leq$0.5 cm at five positions. 

\section{NEW DATA FOR DOUBLE-BETA DECAY OF $^{100}$M\lowercase{o} TO THE FIRST EXCITED 0$^+$ STATE OF $^{100}$R\lowercase{u}}
The improvement by a factor of 2.4 in the determination of the coincidence efficiency for the $2 \nu \beta \beta$ decay apparatus of \cite{Hor} justifies reducing the statistical uncertainty of T$_{1/2}$=[6.0$^{+1.9}_{-1.1}$(stat) $\pm$0.6(syst)] $\times$ 10$^{20}$ y obtained in \cite{Hor} for the half-life time of $^{100}$Mo for the transition to the first excited 0$_1^+$ state in $^{100}$Ru.  Only in the very recent result of the NEMO group of T$_{1/2}$=[5.7$^{+1.3}_{-0.9}$ (stat) $\pm$ 0.8 (syst)]$\times$10$^{20}$ y \cite{Arn} is the statistical uncertainty appreciably smaller and approaches the systematic uncertainty of the experimental data. However, in order for 2$\nu\beta\beta$ decay half-life time measurements to the 0$_1^+$ state to play a decisive role in testing nuclear matrix element calculations, statistical and systematic accuracies each approaching the $\sim$5\% level are needed for $^{100}$Mo.

\begin{figure}
\begin{center}
 \includegraphics[width = 5.25in]{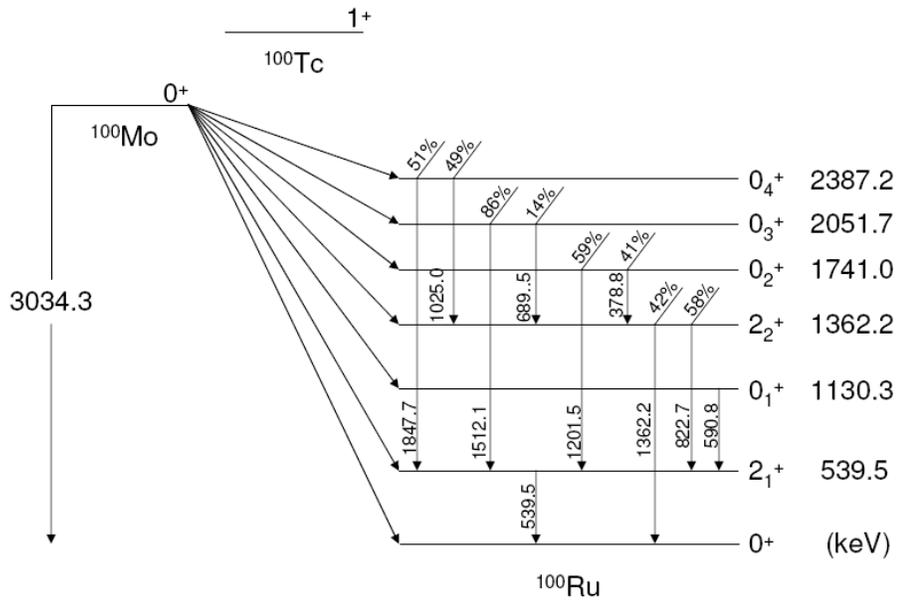}%
 \caption{Level scheme of $^{100}$Ru. Only levels and transitions of interest to the present work are shown.\label{fig:level}}
\end{center}
 \end{figure}

\begin{figure}
\begin{center}
 \includegraphics[width = 5.25in]{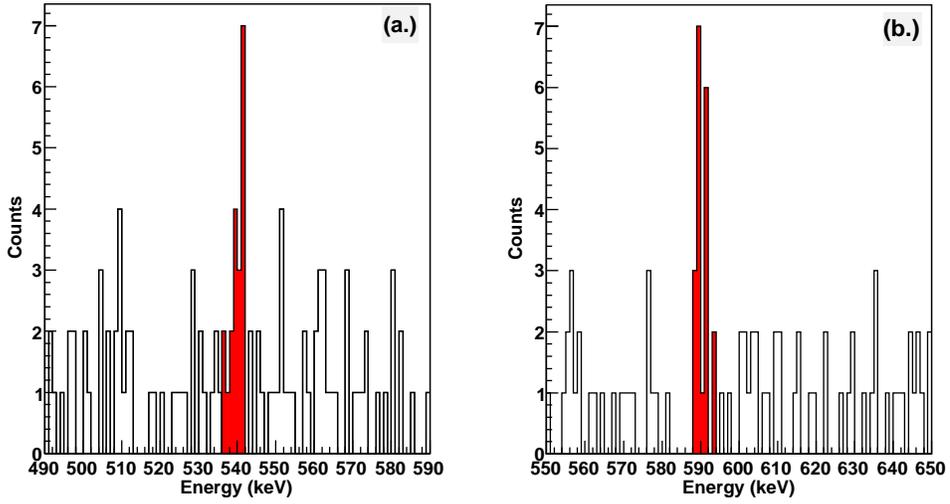}%
 \caption{These spectra show our recent collected events over 450 days.  (a) $\gamma$--ray spectrum in coincidence with the 590.8 $\pm$ 3.0 keV transition in $^{100}$Ru, and (b) $\gamma$--ray spectrum in coincidence with the 539.5 $\pm$ 3.0 keV transition in $^{100}$Ru.\label{fig:events}}
\end{center}
 \end{figure}

\begin{figure}
\begin{center}
 \includegraphics[width = 5.25in]{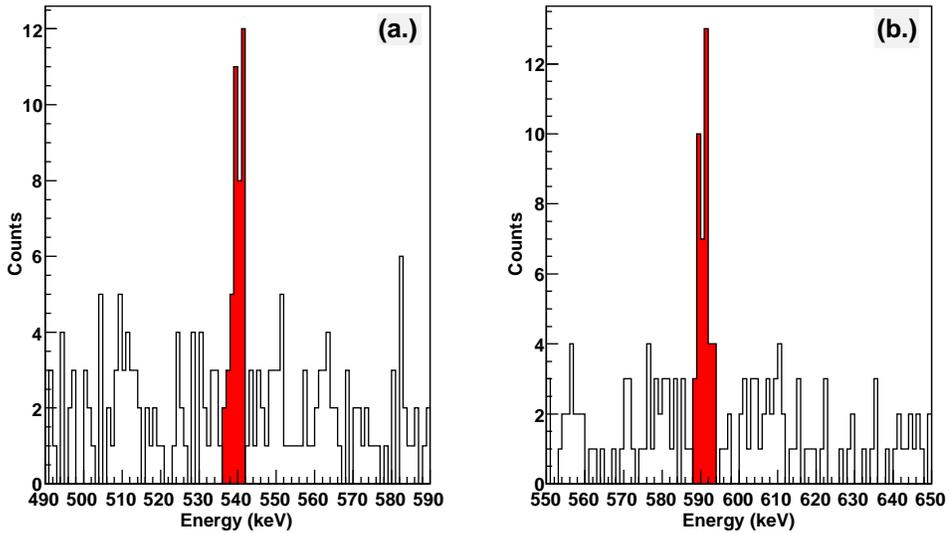}%
 \caption{These spectra show the total collected events with the TUNL-ITEP apparatus over 905 days.  (a) $\gamma$--ray spectrum in coincidence with the 590.8 $\pm$ 3.0 keV transition in $^{100}$Ru, and (b) $\gamma$--ray spectrum in coincidence with the 539.5 $\pm$ 3.0 keV transition in $^{100}$Ru.\label{fig:totalevents}}
\end{center}
 \end{figure}

In an attempt to approach this level of accuracy also for the statistical uncertainty, we report in this section on a new result of our ongoing measurement of T$_{1/2}$ for $^{100}$Mo to the 0$_1^+$ state in $^{100}$Ru. The part of the level scheme of $^{100}$Ru which is of interest to the present work is given in Fig. \ref{fig:level}. Because the approach and analysis procedures are the same as those described in \cite{Hor}, only the essential facts are given here. After 450 days of data taking we recorded 16.0 $\pm$ 4.4 net coincidence events after subtraction of 0.45 $\pm$ 0.16 background events per keV. This result can be compared to the 19.5 $\pm$ 4.7 net events reported in \cite{Hor} for 455 days of data taking. Figure \ref{fig:events} shows the energy spectra for the 539.5 keV and 590.8 keV transitions obtained in the present work, while Fig. \ref{fig:totalevents} shows our spectra combined with those of \cite{Hor}, resulting in a total of 35.5 $\pm$ 6.4 coincidence events.  Using our newly determined coincidence efficiency discussed in Sec. 2 we obtain
 T$_{1/2}$=[6.0$^{+2.3}_{-1.3}$ (stat) $\pm$ 0.3 (syst)]$\times$10$^{20}$ y
for the present data set.  We have no indication that the efficiency of the TUNL 2$\nu\beta\beta$ decay apparatus has changed between the data of \cite{Hor} and the present data were accumulated. Therefore, we used our newly determined coincidence efficiency for the data of \cite{Hor} and obtained
T$_{1/2}$=[5.0$^{+1.6}_{-0.9}$ (stat) $\pm$ 0.3 (syst)]$\times$10$^{20}$ y.  
Combining this result with our present result gives
T$_{1/2}$=[5.5$^{+1.2}_{-0.8}$ (stat) $\pm$ 0.3 (syst)]$\times$10$^{20}$ y.  
This value is in agreement with the recommended value of Barabash \cite{BarCon} for the decay of $^{100}$Mo to the 0$_1^+$ state in $^{100}$Ru of
T$_{1/2}$=6.2$^{+0.9}_{-0.7}$$\times$10$^{20}$ y.  Using the phase-space factor value G$_{2\nu}$=1.64 $\times$ 10$^{-19}$ y$^{-1}$ (for g$_A$ = 1.254) and the measured half-life, one obtains the nuclear matrix element value for the 2$\nu\beta\beta$ transition M$_{2\nu}$(0$^+_1$)=0.105 $\pm$ 0.010 (scaled by the electron rest mass).

Because the technique used in \cite{Hor} and in the present work does not allow a distinction between 0$\nu\beta\beta$ and 2$\nu\beta\beta$ decay, strictly speaking, our result for T$_{1/2}$ is the sum of the 0$\nu\beta\beta$ and 2$\nu\beta\beta$ decay half-life times. However, the experimental limit for the 0$\nu\beta\beta$ decay to the 0$^+$ ground state of $^{100}$Ru is about three orders of magnitude larger \cite{Arn} than the value reported here for T$_{1/2}^{0\nu\beta\beta+2\nu\beta\beta}$.  Therefore, considering the reduced phase space available for the transition to the excited 0$_1^+$ state, it is safe to assume that our result for T$_{1/2}$ refers solely to the 2$\nu\beta\beta$ decay.  Additionally, theoretical estimates for 0$\nu\beta\beta$ decay to this level for a neutrino mass of 1 eV are many orders of magnitude higher:   (7.6 - 14.6) $\times$ 10$^{24}$ y \cite{Sim} and 2.6 $\times$ 10$^{26}$ y \cite{Suh}.

In Ref. \cite{Bar07} the half-life ratio, $R_T = T_{1/2}(0^+_1)/T_{1/2}(0^+_{g.s.})$ was calculated for $^{100}$Mo using the single-state dominance approach of \cite{Dom05} and applying different scenarios for neutrino statistics.  For purely fermionic neutrinos this ratio turned out to be $R_T=61$ while for purely bosonic neutrinos the value $R_T=73$ was calculated. Using the very precise result of T$_{1/2}$=(7.11$\pm$0.54)$\times$10$^{18}$ y from the NEMO-3 experiment \cite{Arn} for the 0$^+$ ground-state transition and our half-life for the 0$^+_1$ excited state transition one obtains the ratio $R_T=77^{+25}_{-16}$. Although this ratio seems to favor the hypothesis of bosonic neutrinos, a more accurate half-life time for the 0$^+_1$ transition is needed to draw any definite conclusions.

\section{NEW DATA FOR DOUBLE-BETA DECAY OF $^{100}$M\lowercase{o} TO HIGHER EXCITED FINAL STATES OF $^{100}$R\lowercase{u}}

Following \cite{Hor} we also determined new limits for the half lives of decays to higher excited states in $^{100}$Ru (see Fig. \ref{fig:level}).  For the 2$\nu\beta\beta$ decay transition to the second excited 2$^+$ state in $^{100}$Ru (2$_2^+$) at 1362.2 keV we searched for the coincident detection of 822 keV and 540 keV $\gamma$ rays. No events of this type were observed.

For the 2$\nu\beta\beta$ decay transition to the second excited 0$^+$ state of $^{100}$Ru (0$_2^+$) at 1741 keV we did not observe any events for the 1201 keV -- 540 keV coincidence pair. For the 379 keV -- 1362 keV coincidence pair one event was observed. This number is consistent with the background yield.

In the case of the transition to the third excited 0+ state (0$_3^+$) at 2051 keV our data show one 1512 keV -- 540 keV event and no 689 keV -- 1362 keV events. The latter event is consistent with the background yield.

Finally, for the transition to the fourth excited 0$^+$ state in $^{100}$Ru (0$_4^+$) at 2387 keV we did not observe any events for the 1848 keV -- 540 keV or the 1025 keV--1362 keV coincidence pairs. 

Using the measured coincidence efficiency of Sec. 2 and correcting it for the energy dependent attenuation in the $^{100}$Mo sample and energy dependent HPGe detector efficiencies, new lower limits for the half--life time of the 2$\nu\beta\beta$ decay transitions discussed above were obtained. We combined these results with those of \cite{Hor} using our newly determined coincidence efficiency and obtained the limits summarized in Table \ref{tab:results}. Basically, the limits were improved by about a factor of two compared to the earlier work \cite{Hor}.  

The new limit for the 2$^+_2$ state transition is of special importance regarding the issue of neutrino statistics. Contrary to the 0$^+_1$ state discussed in Sec. 3, the ratio $R_T$ calculated using excited 2$^+$ states instead of excited 0$^+$ states is very different for bosonic and fermionic neutrinos. According to \cite{Bar07} the ratio $R_T$ decreases for bosonic neutrinos by two orders of magnitude for the first excited 2$^+_1$. Of course, this state is inaccessible with our coincidence technique, but the 2$^+_2$ state certainly is, and it is conceivable that our present limit of T$_{1/2}>$4.4$\times$10$^{21}$ y for this state can be improved considerably in a dedicated experiment.

\section{SUMMARY AND CONCLUSION}

We performed an improved measurement of the coincidence detection efficiency of the TUNL-ITEP double-beta decay apparatus. The efficiency has been determined to an overall accuracy of 5.1\%, providing the basis for determining 2$\nu\beta\beta$ transitions to excited states in daughter nuclei to higher precision than previously possible. We also present new data obtained during a 450 day measurement of T$_{1/2}$ for the 2$\nu\beta\beta$ decay of $^{100}$Mo to the first excited 0$_1^+$ state in $^{100}$Ru. Combining this data set with the previous data of \cite{Hor} resulting in a total accumulation time of 905 days, we obtained the most accurate value for the half-life and NME of this decay.  This half-life and our lower bound for the transition to the 2$^+_2$ state in $^{100}$Ru could be used to set bounds on a speculated bosonic component to neutrino statistics.  

\section{ACKNOWLEDGEMENTS}

This work was supported in part by the U.S. Department of Energy, Office of Nuclear Physics under grant number DE--FG02--97ER41033 and also in part by the U.S. Civilian Research and Development Foundation under grant number RUP1-2892-MO-07.

%\begin{tablehere}
\begin{table}
\caption{Summary of systematic error contributions.}
\begin{tabular}{lc} 
\hline
\hline
Uncertainty Contribution  & \% \\
\hline
Intensity of Calibration Gamma Source & 3\%\\
Energy and Attenuation Correction Factor & 1\%\\
{\it z}-dependence Correction Factor & 1\%\\
Geometry of $^{100}$Mo source & 3\%\\
Non-Symmetrical Efficiency Curve & 2.4\%\\
Dead Time & 0.15\%\\
Uncertainty in $^{102}$Rh half-life & 0.15\%\\
Total & 5.1\%\\
\hline
\hline
\end{tabular}
\label{tab:error}
%\end{tablehere}
\end{table}

%\begin{tablehere}
\begin{table}
\caption{Summary of half--life times and limits for the 2$\nu\beta\beta$ decay of $^{100}$Mo to various excited states of $^{100}$Ru. Limits are given at the 90\% C.L.  The results are based on the combination of the previous data \cite{Hor} and the present data.  See Table II of Hornish {\it et al.} \cite{Hor}}
\begin{tabular}{ccccc} 
\hline
\hline
Transition & Level (keV)  & Q$_{\beta\beta}$ (keV) &  T$_{1/2}^{0\nu+2\nu}$ ($\times$ 10$^{20}$ yr) & T$_{1/2}^{0\nu+2\nu}$ ($\times$ 10$^{20}$ yr) from \cite{Hor}\\
\hline
0$^+\rightarrow$0$^+_1$ & 1130.3 & 1904.0 & 5.5$^{+1.2}_{-0.8}$ (stat) $\pm$ 0.3 (syst) & 6.0$^{+1.9}_{-1.1}$ (stat) $\pm$ 0.6 (syst)\\
0$^+\rightarrow$2$^+_2$ & 1362.2 & 1672.1 & $>$44 & $>$27\\
0$^+\rightarrow$0$^+_2$ & 1741.0 & 1293.3 & $>$48 & $>$28\\
0$^+\rightarrow$0$^+_3$ & 2051.7 & 982.6 & $>$43 & $>$24\\
0$^+\rightarrow$0$^+_4$ & 2387.2 & 647.1 & $>$40 & $>$25\\

%took out 68% CL's
%0$^+\rightarrow$2$^+_2$ & 1362.2 & 1672.1 & $>$89(44) & $>$54(27)\\
%0$^+\rightarrow$0$^+_2$ & 1741.0 & 1293.3 & $>$94(48) & $>$55(28)\\
%0$^+\rightarrow$0$^+_3$ & 2051.7 & 982.6 & $>$76(43) & $>$42(24)\\
%0$^+\rightarrow$0$^+_4$ & 2387.2 & 647.1 & $>$79(40) & $>$49(25)\\
\hline
\hline
\end{tabular}
\label{tab:results}
%\end{tablehere}
\end{table}
% The Appendices part is started with the command \appendix;
% appendix sections are then done as normal sections
% \appendix

% \section{}
% \label{}

\end{document}